\documentclass[aps,twocolumn,prb,showpacs,preprintnumbers,amsmath,amssymb]{revtex4}
\newcommand{\lsim} 
 {\ \raise.35ex\hbox{$<$}\kern-0.75em\lower.5ex\hbox{$\sim$}\ }
\newcommand{\gsim}
 {\ \raise.35ex\hbox{$>$}\kern-0.75em\lower.5ex\hbox{$\sim$}\ }
\newcommand{\bras}[1]{\langle#1|}
\newcommand{\kets}[1]{|#1\rangle}
\newcommand{\brasd}[1]{\langle\!\langle#1|}
\newcommand{\ketsd}[1]{|#1\rangle\!\rangle}

\newcommand{\means}[1]{\langle#1\rangle}

\def\journal #1#2#3#4{#1 {\bf #2}, #3 (#4)}
\def\PR{Phys.\ Rev.}

\def\PRB{Phys.\ Rev.\ B}
\def\PRL{Phys.\ Rev.\ Lett.}

\def\JAP{J.\ Appl.\ Phys.}

\def\JPSJ{J.\ Phys.\ Soc.\ Jpn.}

\usepackage{graphicx}
\usepackage{dcolumn}
\usepackage{bm}
\usepackage{amsmath}
\usepackage{times}
\usepackage[usenames]{color}

\usepackage{ulem}

\begin{document}
%\draft
%\preprint{?????}
\title{Vibronic Excitation Dynamics in Orbitally Degenerate Correlated Electron System}
\author{Joji~Nasu$^\ast$ and Sumio~Ishihara} 
 \affiliation{Department of Physics, Tohoku University, Sendai 980-8578, Japan} 
\date{\today}
\begin{abstract}  
Orbital-lattice coupled excitation dynamics in orbitally degenerate correlated systems are examined. We present a theoretical framework, where both local vibronic excitations and superexchange-type inter-site interactions are dealt with on an equal footing. We generalize the spin-wave approximation so as to take local vibronic states into account. Present method is valid from weak to strong Jahn-Teller coupling magnitudes. Two characteristic excitation modes coexist; a low-energy dispersive mode and high-energy multi-peak mode. These are identified as a collective vibronic mode, and Flanck-Condon excitations in a single Jahn-Teller center modified by the inter-site interactions, respectively. Present formalism covers vibronic dynamics in several orbital-lattice coupled systems. 
\end{abstract}

%75.25.Dk 	Orbital, charge, and other orders, including coupling of these orders
%75.30.Et 	Exchange and superexchange interactions (see also 71.70.Gm Exchange interactions)
%75.47.Lx 	Magnetic oxides

\maketitle
%\narrowtext

%--- title ---

%--- author ---

%
%
%--- address ---

%
%--- date ---

%   Abstract
%-----------------------------------------------------------

%-----------------------------------------------------------

%\section{Introduction\label{sec:intro}}
%%%%%%%%%%%%%%%%%%%%%%%%%%%%%%%%%%%%%%%%%%

\section{Introduction}

Orbital degree of freedom of an electron represents a directional aspect of electronic wave function. It is widely recognized that the orbital degree of freedom influences significantly magnetic, optical, and structural properties in correlated electron materials.~\cite{Tokura2000, Maekawa2004} A macroscopic symmetry breaking of a degenerate orbital wave function, termed an orbital order, is often seen in several transition-metal compounds, rare-earth magnets, as well as molecular solids. A long-range orbital order is generally accompanied with a macroscopic lattice distortion which is compatible to a shape of the electronic wave function. This is caused by an orbital-lattice interaction known as the Jahn-Teller (JT) effect in a single molecule. 

A collective orbital excitation in an orbital ordered state is termed ``orbital wave'' and its quantized object is termed ``orbiton''. This is an analogous to spin-wave excitation in a magnetically ordered state. Several experimental observations of orbiton by optical and Raman spectroscopies,~\cite{Saitoh2001,Miyasaka2005,Ulrich2006,Sugai2006,Benckiser2008}  x-ray scatterings~\cite{Inami2003,Tanaka2004,Ulrich2009,Ishii2011,Schlappa2012} and other experimental probes, as well as theoretical supports~\cite{Inoue1997,Ishihara2004,Ishihara2005,Haverkort2010,Ament2011} have been reported so far. Nonetheless, characteristics of orbiton have not been revealed yet. This might be attributed to a fact that coupling between orbiton and lattice is not negligibly small, and an experimental assignment of orbiton is not so simple in comparison with that of magnon.

The first theoretical examination of orbital wave was done by Cyrot and Lyon-Caen,~\cite{Cyrot1975} and Komarov {\it et al.},~\cite{Komarov1975} where purely electronic orbital excitations as well as spin-orbital coupled excitations were examined based on a correlated electron model. More realistic calculations of orbital wave were performed by one of the present authors and coworkers in Refs.~\onlinecite{Ishihara1996, Ishihara1997, Ishihara2000}, where the lattice distortion is interpreted to be frozen. The adiabatic frozen-lattice treatment~\cite{Bala2000} is justified in the limiting case where the orbital excitation energy is much higher than the phonon energy. A weak coupling approach for the JT effect was adopted in Ref.~\onlinecite{VandenBrink2001} where an anticrossing-type mixing between orbiton and phonon branches occurs. Similar linear coupling between orbiton and phonon modes are examined in TmVO$_4$ and related materials.~\cite{Gehring1975,Kaplan1995} 
On the other side, vibronic excitations in a single JT center have been examined intensively.~\cite{Bersuker2006} In particular, local vibronic excitations in an orbital ordered state were studied in Ref.~\onlinecite{Allen1999}, where multi-peak vibrational excitations with a broad envelop appear due to the Franck-Condon transitions. 

Purpose in this paper is to present a theoretical framework of vibronic excitations in orbitally degenerate correlated electron system; both the local vibronic excitations and the superexchange (SE)-type inter-site interaction between orbitals are taken into account on an equal footing. We set up a model which consists of the SE interactions, the on-site JT coupling and the local lattice vibration. A generalized spin-wave approach where the local vibronic states are fully taken into account is presented. Two characteristic excitation modes coexist; a low-energy dispersive vibronic mode interpreted as a renormalized ``orbiton'', and high-energy multi-peaks originating from the Franck-Condon excitation in a single JT center modified by the SE interaction. The present formalism does not only cover orbitally degenerate systems from weak to strong JT couplings, but also is applicable to several orbital-lattice coupled models.

In Sec.~\ref{sec:model}, we introduce a model Hamiltonian for an orbital-lattice coupled system. In Sec.~\ref{sec:formulation-method}, a generalized spin-wave approximation for vibronic excitations is presented. Before showing the detailed numerical calculations, results obtained by the present theory are compared with the results by the exact diagonalization method in Sec.~\ref{sec:valid-pres-meth}. The main part in this paper is Sec.~\ref{sec:main}, where the detailed energy and momentum dependences of the vibronic excitations are presented. In Sec.~\ref{sec:low-energy-excit}, we focus on the low-energy excitation modes corresponding to the collective vibronic modes. In Sec.~\ref{sec:comp-betw-cases}, results in the present $E\otimes e$ system are compared with those in the $E\otimes b_1$ system. Section~\ref{sec:discussion-summary} is devoted to discussion and summary.

\section{Model}\label{sec:model}

In order to address an issue for the coupling between the orbital excitation and the lattice dynamics, we introduce an $E \otimes e$ JT center at each lattice site and the SE-type interactions between the nearest-neighbor (NN) $e_g$ orbitals. We adopt the following orbital-lattice coupled Hamiltonian, 
\begin{align}
 {\cal H}={\cal H}_J+{\cal H}_{\rm JT}.
\label{eq:3}
\end{align}
The first term, ${\cal H}_J$, represents the SE interactions and the second term, ${\cal H}_{\rm JT}$, is for the local lattice vibration and the JT coupling. One of the prototypical SE interaction in an orbitally degenerate magnet is the Kugel-Khomskii type spin-orbital Hamiltonian~\cite{Kugel1972,Kugel1974} which is derived from the multi-orbital Hubbard model. Here, we focus on the orbital degree of freedom in the SE-type interaction, and consider the following spin-less orbital-only model,
\begin{align}
 {\cal H}_J&=-\sum_{\langle ij \rangle}\left (J_z T_i^z T_j^z+J_x T_i^x T_j^x \right),
 \label{eq:12}
\end{align}
where NN $ij$ sites are represented by $\langle ij \rangle$. The doubly-degenerate orbitals are described by the pseudo-spin (PS) operator defined by $\bm{T}_i=\frac{1}{2}\sum_{\gamma \gamma'} d_{i\gamma}^\dagger \bm{\sigma}_{\gamma\gamma'} d_{i\gamma'}$, where $d_{i\gamma}$ is an annihilation operator for a spin-less fermion with orbital $\gamma$ at site $i$, and ${\bm \sigma}$ are the Pauli matrices. The eigen state of $T^z$ with the eigen value of $+1/2$ $(-1/2)$ corresponds to a state where the $d_{3z^2-r^2}$ $(d_{x^2-y^2})$ orbital is occupied by an electron. The exchange constants, $J_z$ and $J_x$, are set to be positive. Present formulation is able to be generalized easily to models where other terms for the PS interactions, such as $T_i^zT_j^x$, and the spin degree of freedom, are taken into account. This will be discussed in Sec.~\ref{sec:discussion-summary}. 

The second term of the Hamiltonian is given by 
\begin{align}
  {\cal H}_{\rm JT}=\sum_i {\cal H}_{i}^{\rm JT}, 
\end{align}
with 
\begin{align}
 {\cal H}_{i}^{\rm JT}&=-\frac{1}{2M}\left(\frac{\partial^2}{\partial Q_{iu}^2}+\frac{\partial^2}{\partial Q_{iv}^2}\right)+\frac{M\omega_0^2}{2}(Q_{iu}^2+Q_{iv}^2)\nonumber\\& \ \ \ \ \ 
+2A(-T_{i}^z Q_{iu}+T_{i}^x Q_{iv}), \label{eq:2}
\end{align}
where $Q_{iu}$ and $Q_{iv}$ represent the two vibrational modes at the $i$-th JT center with the $E_g$ symmetry. The first two terms describe the harmonic vibrations with frequency $\omega_0$ and a reduced mass $M$, and the third term describes the linear JT coupling with a coupling constant $A(>0)$. 
For convenience, we introduce the phonon coordinates for lattice vibrations at $A=0$ as follows. The creation and annihilation operators for the mode $\gamma(=u, v)$ phonons are defined by $b_i^{\gamma\dagger}=[-l_0\partial/(\partial Q_{i\gamma})+Q_{i \gamma}/l_0]/\sqrt{2}$ and $b_i^{\gamma}=[l_0\partial/(\partial Q_\gamma)+Q_{ i\gamma}/l_0]/\sqrt{2}$ with $l_0=(M\omega_0)^{-1/2}$, respectively. Then, this term of the Hamiltonian is rewritten as 
\begin{align}
 {\cal H}_{i}^{\rm JT}= \omega_0 \sum_{\gamma}   b_i^{\gamma\dagger}b_i^{\gamma} 
 -gT_i^z(b_i^{u\dagger}+b_i^u)+gT_i^x(b_i^{v\dagger}+b_i^v) ,
\label{eq:10}
\end{align}
where the coupling constant is defined by $g=\sqrt{2} A l_0$. We neglect the higher-order JT coupling, the anharmonic lattice potential, and the cooperative JT effect, for simplicity, although the present formulation is generalized easily to include these effects (see Sec.~\ref{sec:discussion-summary}). 

It is worth noting that ${\cal H}_{i}^{\rm JT}$ is invariant under the simultaneous infinitesimal rotations of PS and ${\bm Q}=(Q_u, Q_v)$ given by $(T_i^z,T_i^x)\rightarrow (T_i^z,T_i^x)+\varepsilon (T_i^x,-T_i^z)$ and $(Q_{iu}, Q_{iv})\rightarrow (Q_{iu}, Q_{iv})+\varepsilon (-Q_{iv}, Q_{iu})$ where $\varepsilon$ is an infinitesimal constant. Therefore, a gapless Goldstone mode~\cite{Sarfatt1964} exists in the case of $J_x=J_z$. 

\section{Formulation}\label{sec:formulation-method}

We present a formulation based on the generalized spin-wave (SW) approximation, where the local vibronic states are fully taken into account. We show later that this is valid from weak to strong JT coupling regimes. A relation of the present formalism to the random-phase approximation (RPA) is given in Appendix~\ref{sec:equiv-with-rand}. Similar formalisms were presented in Refs.~\onlinecite{Papanicolaou1988,Onufrieva1985,Shiina2003,Kusunose2001,Joshi1998}. 

We assume that $J_z\geq J_x$ in ${\cal H}_J$ and the uniform orbital order for $T^z$ in the ground state, without loss of generality. The $z$ component of the PS operator is decomposed into the ordered moment and fluctuation as 
$T_i^z=\means{T^z}+\delta T_i^z$  
where $\means{\cdots}$ denotes the expectation value in the ground state. The Hamiltonian in Eq.~(\ref{eq:3}) is rewritten as 
\begin{align}
 {\cal H}=-\sum_{\means{ij}}(J_z \delta T_i^z \delta T_j^z+J_x T_i^x T_j^x)
+\sum_i {\cal H}_i^{\rm MF} .
\label{eq:5}
\end{align}
We define the on-site term 
\begin{align}
{\cal H}_i^{\rm MF}=-h_{\rm MF}T_i^z +{\cal H}_i^{\rm JT} , 
\label{eq:hmf}
\end{align}
with  
\begin{align}
h_{\rm MF}=zJ_z\means{T^z} , 
\label{eq:hmf2}
\end{align}
where $z$ is a coordination number. 
 
The ordered moment $\means{T^z}$ is determined by the following way. The local Hamiltonian ${\cal H}_i^{\rm MF}$ is diagonalized numerically under a given $\means{T^z}$. The eigen states $\{\kets{\Phi_{n}}\}$ and the eigen energies $\{E_n \}$ are obtained up to ${\cal N}(\ge n)$ where the phonon-number is restricted to be less than $N_{ph}$ at each site. In the present numerical calculations, we chose $N_{ph}=80$, which is enough to examine excitations of the present interest. The ordered moment is calculated by the ground-state wave-function $\kets{\Phi_0}$ as $\means{T^z}=\bras{\Phi_0}T^z\kets{\Phi_0}$. This procedure is repeated until $\means{T^z}$ converges. It is noted that ${\cal H}^{\rm MF}_i$ commutes with the parity operator ${\cal P}_i=2T^z_i e^{i\pi b^{v\dagger}_i b^v_i }$, and the eigen states are classified by the eigen values of ${\cal P}_i$, i.e. ${\cal P}_i \kets{\Phi_n}=p_n \kets{\Phi_n}$. When $p_n=1$ $(p_n=-1)$, a parity of $\kets{\Phi_n}$ is identified as ``even'' (``odd''). A parity of the ground state is even. 

By using the calculated eigen states, the PS operators are expanded by the projection operators (X-operators) as 
\begin{align}
T_i^x&=\sum_{m, n=0}^{\cal N} (T^x)_{mn} X_i^{mn},    
\end{align}
and 
\begin{align}
\delta T_i^z&=\sum_{m, n=0}^{\cal N} (\delta T^z)_{mn} X_i^{mn} , 
\end{align}
where 
$X_i^{mn}=\kets{\Phi_{im}}\bras{\Phi_{in}}$, 
$(T^x)_{mn}=\bras{\Phi_{im}}T_i^x\kets{\Phi_{in}}$ and $(\delta T^z)_{mn}=\bras{\Phi_{im}}\delta T_i^z\kets{\Phi_{in}}$. 
By applying the generalized Holstein-Primakoff transformation,~\cite{Papanicolaou1988,Onufrieva1985,Shiina2003,Kusunose2001,Joshi1998} the projection operators are represented by the boson operators $a_{im}$ as
\begin{align}
X_i^{mn}&=a_{in}^\dagger a_{im} , 
\end{align}
for $n, m \geq 1$, 
\begin{align} 
X_i^{n0}= a_{in}^\dagger \left(M-\sum_{m=1}^{\cal N} a_{im}^\dagger a_{im}\right)^{1/2}, 
\end{align}
for $n \geq 1$, 
\begin{align}
X_i^{00}=M-\sum_{n=1}^{\cal N} a_{in}^\dagger a_{in}, 
\end{align}
and $X_i^{0n}=(X_i^{n0})^\dagger$. A constraint $M \equiv X_i^{00}+\sum_{n=1}^{\cal N} a_{in}^\dagger a_{in}=1$ is imposed. The commutation relations for the projection operators, $[X_i^{mn}, X_j^{m'n'}]=\delta _{ij}  ( X_i^{mn'}\delta_{nm'}-X_i^{m'n}\delta_{n'm} )$, are derived by the constraint and commutation relations for $a_m$ and $a_m^\dagger$. The SU(2) commutation relations for the PS operators are insured, when $\cal N$ is taken to be infinity. By the $1/M$ expansion up to ${\cal O}(M^{1/2})$, we have 
\begin{align}
T_i^x&=M^{1/2}\sum_{m=1}^{\cal N}v_m^x (a_{im}+a_{im}^\dagger) ,  
\label{eq:tx}
\end{align}
\begin{align}
\delta T_i^z&=M^{1/2}\sum_{m=1}^{\cal N}v_m^z (a_{im}+a_{im}^\dagger) , 
\label{eq:tz}
\end{align}
where $v_m^x=\bras{\Phi_0}T^x\kets{\Phi_m}$ and $v_m^z=\bras{\Phi_0}\delta T^z\kets{\Phi_m}$. 

Then, the Hamiltonian is given as a bilinear form for the boson operators as 
\begin{align}
{\cal H}&=\sum_{\bm{q}}\sum_{m,n}^{( {\rm even})}\bigl[(\Delta E_n \delta_{mn}-z \gamma_{\bm{q}}J_z v_m^z v_n^z)a_{\bm{q}m}^\dagger a_{\bm{q}n}\nonumber\\
&-\frac{z\gamma_{\bm{q}}J_z}{2} v_m^z v_n^z(a_{\bm{q}m}^\dagger a_{-\bm{q}n}^\dagger+h.c)\bigr]\nonumber\\
&+\sum_{\bm{q}}\sum_{m,n}^{( {\rm odd})}\bigl[(\Delta E_n \delta_{mn}-z\gamma_{\bm{q}}J_x v_m^x v_n^x)a_{\bm{q}m}^\dagger a_{\bm{q}n}
\nonumber\\
&-\frac{z\gamma_{\bm{q}}J_x}{2} v_m^x v_n^x(a_{\bm{q}m}^\dagger a_{-\bm{q}n}^\dagger+h.c)\bigr],\label{eq:1}
\end{align}
where $\Delta E_n=E_n-E_0$, $a_{\bm{q}n}$ is the Fourier transform of $a_{in}$, and $\gamma_{\bm{q}}=z^{-1}\sum_{\bm{\rho}}e^{i\bm{q}\cdot\bm{\rho}}$ is a form factor where summations are taken for the NN sites. A symbol $\sum_{m,n}^{( {\rm even \ (odd)})}$ represents a summation for the even (odd) parity states. This originates from $v_n^x=0$ ($v_n^z=0$) for the even (odd) parity states due to the relations ${\cal P}T^z {\cal P}=T^z$ and ${\cal P}T^x {\cal P}=-T^x$.
The Hamiltonian in Eq.~(\ref{eq:1}) is diagonalized by the generalized Bogoliubov transformation~\cite{Colpa1978} as 
\begin{align}
 {\cal H}=\sum_{\bm{q}}\sum_{\eta}\Omega_{\bm{q}\eta}\alpha_{\bm{q}\eta}^\dagger\alpha_{\bm{q}\eta}+{\rm const.},
 \label{eq:omega}
\end{align}
where $\alpha_{\bm{q}\eta}$ is a boson operator given by a linear combination of sets of $\{a_{\bm{q}m}\}$ and $\{a_{-\bm{q}m}^\dagger\}$, and $\Omega_{\bm{q}\eta}$ is the eigen energy. The ground state of ${\cal H}$, termed $\ketsd{0}$, is defined as a vacuum of $\alpha_{\bm{q}\eta}$ for any ${\bm q}$ and $\eta$.

In the present formalism, the PS dynamical susceptibility is given by 
\begin{align}
 \chi^{ll}(\bm{q},\omega)&=-i\int_{0}^\infty dt\brasd{0}[\tilde T_{-\bm{q}}^l(t),\tilde T_{\bm{q}}^l]\ketsd{0}e^{i\omega t-\epsilon t}\nonumber\\
&=\int_{-\infty}^\infty dE\frac{S^{ll}({\bm q}, E)}{\omega-E+i\epsilon},\label{eq:19}
\end{align}
where ${\tilde T_{\bm{q}}^l}=(\delta T^z_{\bm{q}}, T^x_{\bm{q}})$ for $l=(z, x)$, $\epsilon$ is a positive infinitesimal constant, and ${\hat O}(t)=e^{i{\cal H}t} {\hat O}e^{-i{\cal H}t}$ is the Heisenberg representation for the operator $\hat O$. The spectral function is straightforwardly calculated as 
\begin{align}
 S^{ll}({\bm q}, E)&=\sum_{mn\eta}v_m^l v_n^l\brasd{0}(a_{-\bm{q}m}+a_{\bm{q}m}^\dagger)\ketsd{\bm{q},\eta}\nonumber\\
&\ \ \times\brasd{\bm{q},\eta}(a_{\bm{q}n}+a_{-\bm{q}n}^\dagger)\ketsd{0}\delta(E-\Omega_{\bm{q}\eta}), \label{eq:20}
\end{align}
where $\ketsd{\bm{q},\eta} \equiv \alpha_{\bm{q}\eta}^\dagger\ketsd{0}$.
The retarded Green's functions for phonons are defined as
\begin{align}
 D^\gamma({\bm q}, \omega)=-i\int_{0}^\infty dt\brasd{0}[{\tilde b}_{-\bm{q}}^\gamma(t), {\tilde b}_{\bm{q}}^{\gamma\dagger}]\ketsd{0}e^{i\omega t-\epsilon t},
\end{align}
where ${\tilde b}_{\bm{q}}^{\gamma}$ is a Fourier transform of ${\tilde b}_{i}^{\gamma}$, which is defined by ${\tilde b}_i^\gamma=(b_{i}^u+\frac{g}{\omega_0}\means{T_i^z}, b_{i}^v)$ for $\gamma=(u, v)$. This is calculated by~\cite{Koller2004}
\begin{align}
  D^\gamma (\bm{q},\omega)&=D_0(\omega)+g^2[D_0(\omega)]^2\chi^{ll}(\bm{q},\omega), 
\end{align}
where $l=z (x)$ for $\gamma=u (v)$, and $D_0(\omega)=1/(\omega-\omega_0+i\epsilon)$ is the bare phonon Green's function.

\begin{figure}[t]
\includegraphics[width=0.8\columnwidth,clip]{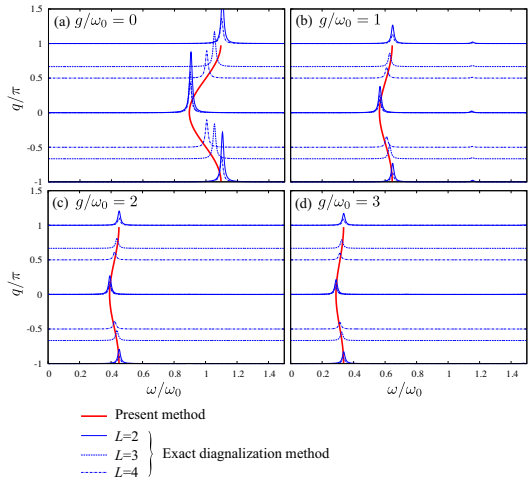}
\caption{(Color online) 
Orbital excitation spectra calculated by the present generalized spin-wave method and the ED method. Red lines are for the poles of ${\rm Im} \chi^{xx}(q, \omega)$, i.e. $\Omega_{\bm{q} \eta}$ defined in Eq.~(\ref{eq:omega}). Blue lines represent $(-1/\pi) {\rm Im} \chi^{xx}({\bm q}, \omega)$ calculated by the ED method. An external field term is added to the Hamiltonian in Eq.~(\ref{eq:3}). In the ED method, cluster size is chosen to be $L=2, 3$ and 4, and an infinitesimal constant as a damping factor of the spectra is chosen to be $\epsilon/\omega_0=0.01$. Parameter values are chosen to be $J_z=0$, $J_x/\omega_0=0.2$, and $h/\omega_0=1$.
}
\label{fig:ED}
\end{figure}
\section{Comparison with Exact Diagonalization Method}
\label{sec:valid-pres-meth}

Before showing detailed results, we compare the numerical results obtained by the present method and the exact diagonalization (ED) method in finite cluster systems, to show validity of the present method. In order to avoid finite size effects in the ED method, we add an external-field term, $-h \sum_i T_i^z$, to the Hamiltonian in Eq.~(\ref{eq:3}), by which  the excitation becomes gapful. In the ED method, Hamiltonian in Eq.~(\ref{eq:3}) plus the external-field term is solved by the Lanczos method, and one-dimensional clusters with a periodic boundary condition are adopted, for simplicity. The parameter values are chosen to be $J_z=0,\ J_x/\omega_0=0.2$, and $h/\omega_0=1$. The Hilbert space is restricted so that the number of the phonons is less than 16 at each site.

In Fig.~\ref{fig:ED}, the PS dynamical susceptibilities calculated by the two methods are compared with each other. We focus on low energy excitations up to $\omega/\omega_0 =1.5$, corresponding to the upper band edge of the collective vibronic excitation, as explained later. It is shown that excitation energies $\Omega_{\bm{q}\eta}$ calculated by Eq.~(\ref{eq:omega}) well reproduce dominant peaks in $(-1/\pi){\rm Im}\chi^{xx}({\bm q}, \omega)$ calculated by the ED method. Good agreements between the two results are seen from the no-coupling case ($g=0$) to the strong coupling case ($g/\omega_0=3$). 

\section{Orbital-Lattice Coupled Vibronic Excitations}
\label{sec:main}
\subsection{Local Vibronic Excitation}
\label{sec:local-excit-struct}
\begin{figure}[t]
\includegraphics[width=0.8\columnwidth,clip]{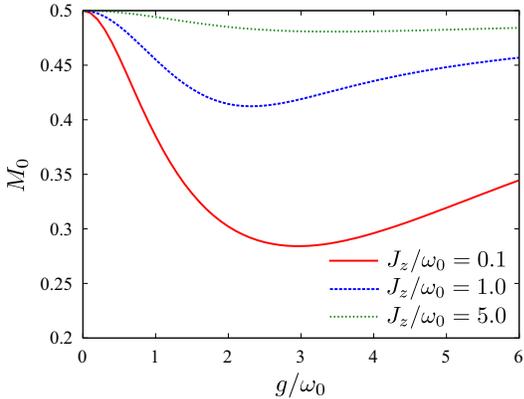}
\caption{(Color online) Ordered moment $M_0=\means{T^z}$ for several $J_z/\omega_0$. A coordination number is chosen to be $z=2$.
}
\label{fig:moment}
\end{figure}
\begin{figure}[t]
\includegraphics[width=0.8\columnwidth,clip]{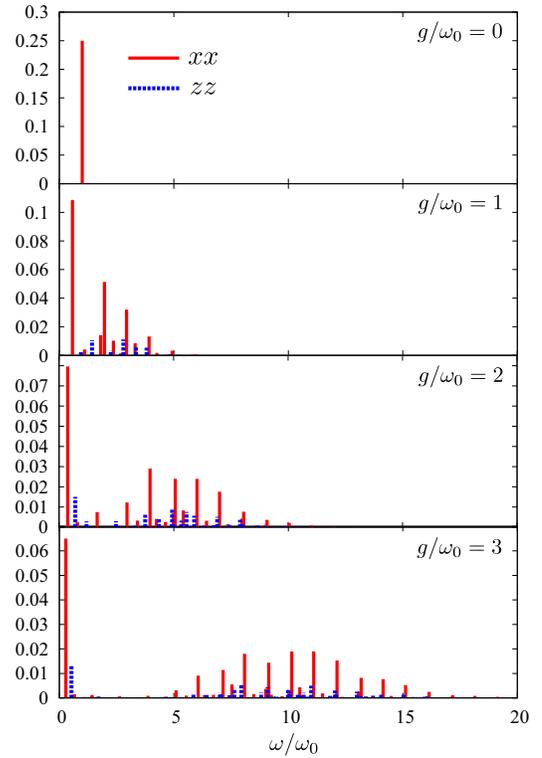}
\caption{(Color online) 
Imaginary parts of the dynamical susceptibilities for several $g/\omega_0$. Red and blue lines represent spectral weights at poles of $(-1/\pi)N^{-1}\sum_{\bm q} {\rm Im} \chi^{xx}({\bm q}, \omega)$ and $(-1/\pi)N^{-1}\sum_{\bm q} {\rm Im} \chi^{zz}({\bm q}, \omega)$, respectively. Parameter values are chosen to be $J_z/\omega_0=1$ and $z=2$.
}
\label{fig:local}
\end{figure}

We start from the results in the local Hamiltonian ${\cal H}_i^{\rm MF}$ defined in Eq.~(\ref{eq:hmf}). The results where the inter-site interaction effects are taken into account are presented in the next subsection.

First, we show the orbital ordered moment $M_0 \equiv \means{T^z}$ as a function of the JT coupling constant in Fig.~\ref{fig:moment}. Non-monotonic behaviors as functions of $g$ are shown. For small $g$, a reduction of $M_0$ with increasing $g$ reflects a suppression of the long-range order due to the vibronic motion. On the other hand, for large $g$, a reduction of $M_0$ from 1/2 decreases with $g$, since the kinetic energy of the lattice vibration is proportional to $1/g^2$ [see the first term in Eq.~(\ref{eq:7})]. A $J_z$ dependence of $M_0$ implies a competition between the vibroic motion and the inter-site SE interaction; a large SE interaction suppresses a reduction of the ordered moment due to the vibroic motion. 

\begin{figure}[t]
\includegraphics[width=\columnwidth,clip]{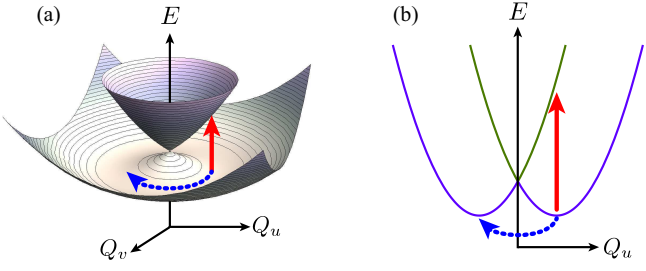}
\caption{(Color online) (a) Adiabatic-potential planes in  the $E\otimes e$ system, and (b) those in the $E\otimes b_1$ system. Red and blue arrows represent excitation between the two adiabatic-potential planes and that in the lowest adiabatic-potential plane.
}
\label{fig:adiabatic_plane}
\end{figure}

Local vibronic excitation spectra, defined by $-(\pi N)^{-1}\sum_{\bm q} {\rm Im} \chi^{ll}({\bm q}, \omega)$, are presented in Fig.~\ref{fig:local}. Here, $N$ represents the number of the lattice sites. At $g=0$, a single peak appears at $\omega=\omega_0$ in the $xx$-component and no finite intensity in the $zz$-component. In finite $g$, two-kind excitations appear; a sharp low-energy peak at a little below $\omega_0$ and high-energy multi-peaks with a Gaussian-like envelope. A center of the envelope is located around $g^2/\omega_0+J_z$. A low energy peak is attributed to the collective vibronic excitation mode of our main interest, and will be examined in more detail in Sec.~\ref{sec:low-energy-excit}. High-energy multi-peaks are attributed to the Frank-Condon excitations from the lower adiabatic-potential plane to the higher plane~\cite{Bersuker2006,Allen1999} A center of the multi-peak structure is located around $g^2/\omega_0+J_z$, i.e. a sum of a separation between the higher and lower adiabatic-potentials and a diagonal component of the SE interaction energy. Schematic pictures for the adiabatic potential planes and transitions in the $Q_u$-$Q_v$ plane are shown in Fig.~\ref{fig:adiabatic_plane}(a).

\subsection{Inter-site Interaction Effect}
\label{sec:effect-inter-site}
\begin{figure}[t]
\includegraphics[width=\columnwidth,clip]{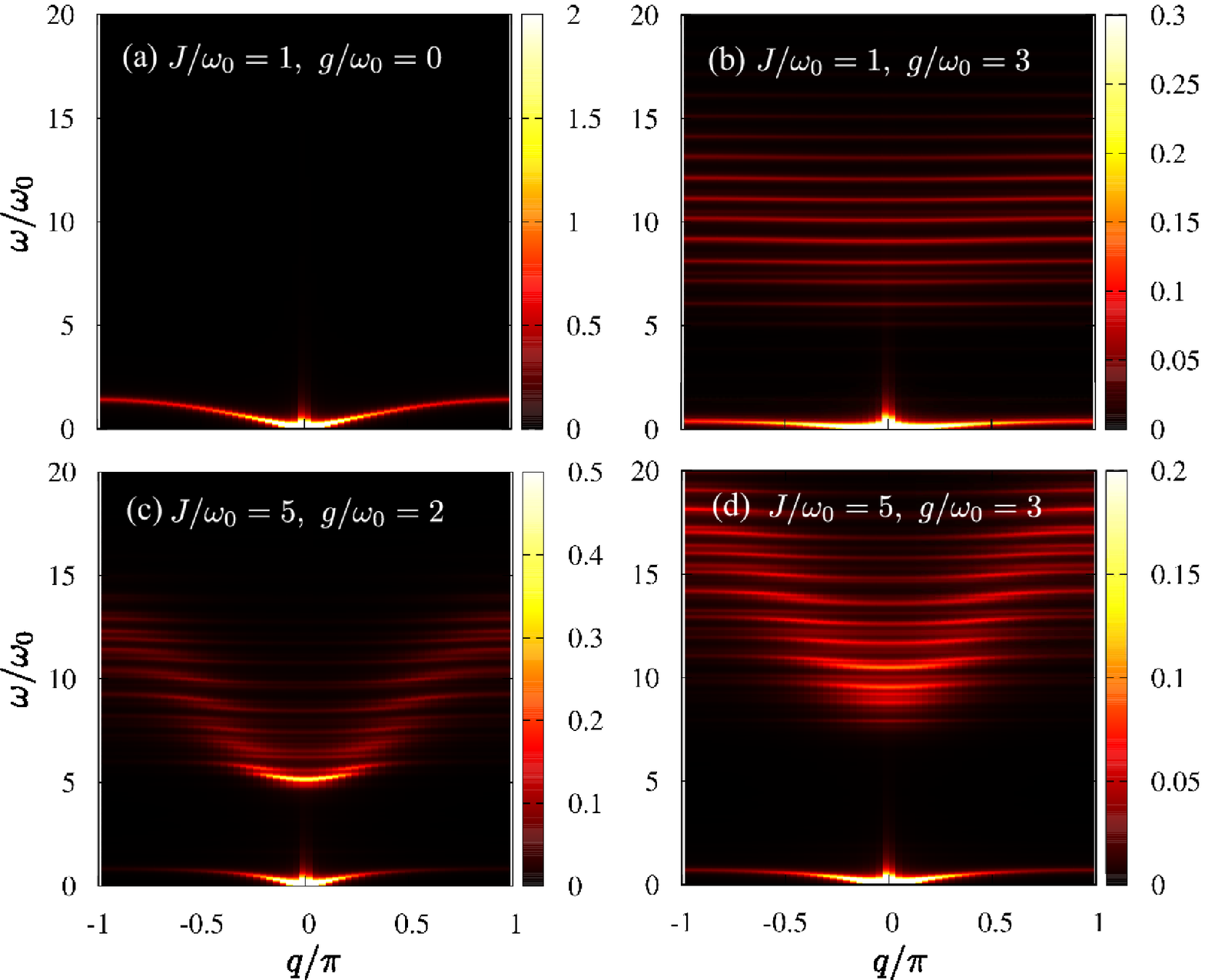}
\caption{(Color online) 
Contour maps of imaginary parts of the dynamical susceptibilities. Colors represent spectral weights $(-1/\pi) {\rm Im} \chi^{xx}({\bm q}, \omega)$. Parameter values are chosen to be (a) $(J/\omega_0, g/\omega_0)=(1,0)$, (b) $(1,3)$, (c) $(5, 2)$, and (d) $(5, 3)$. An infinitesimal constant as a damping factor of the spectra is chosen to be $\epsilon/\omega_0=0.1$.
}
\label{fig:disp}
\end{figure}
\begin{figure}[t]
\includegraphics[width=\columnwidth,clip]{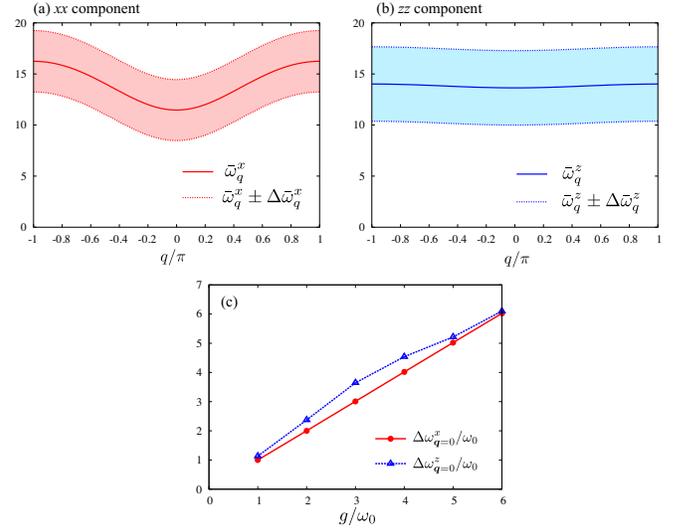}
\caption{(Color online) 
Momentum dependences of the first moment ($\bar{\omega}_{\bm q}^l$) and a square of the second moment ($\Delta \omega_{\bm q}^l$) for the high energy multi-peaks. (a) and (b) are for the $xx$- and $zz$-components, respectively. Bold and dotted lines represent $\bar{\omega}_{\bm q}^l$ and $\bar{\omega}_{\bm q}^l \pm \Delta \omega_{\bm q}^l$, respectively. Parameter values are chosen to be $(J/\omega_0, g/\omega_0)=(5,3)$. (c) Square roots of the second moment ($\Delta \omega_{\bm q}^l$) at ${\bm q}=0$ of the high energy multi-peaks as functions of the JT coupling. We chose $J/\omega_0=5$. 
}
\label{fig:disp_width}
\end{figure}
\begin{figure}[t]
\includegraphics[width=\columnwidth,clip]{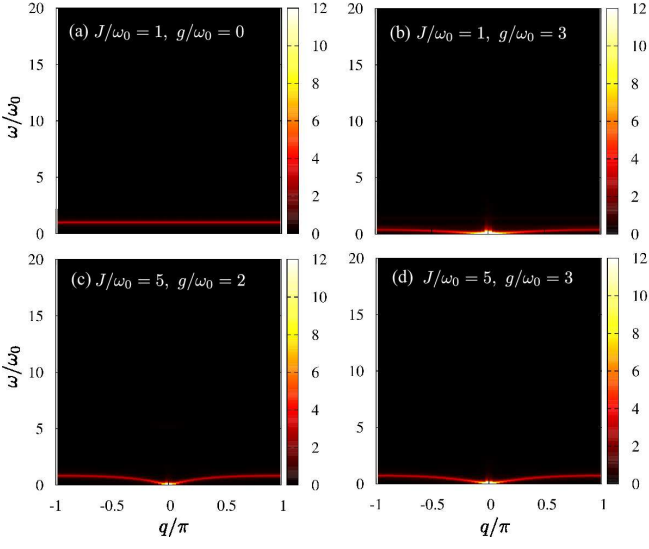}
\caption{(Color online)  
Contour maps of imaginary parts of the phonon Green's function. Colors represent spectral weights $(-1/\pi) {\rm Im} D^{v}({\bm q}, \omega)$. Parameter values are chosen to be (a) $(J/\omega_0, g/\omega_0)=(1,0)$, (b) $(1,3)$, (c) $(5, 2)$, and (d) $(5, 3)$. An infinitesimal constant as a damping factor of the spectra is chosen to be $\epsilon/\omega_0=0.1$.
}
\label{fig:disp_D}
\end{figure}

Numerical results for the Hamiltonian ${\cal H}={\cal H}_{\rm JT}+{\cal H}_J$ are presented. For simplicity, a one-dimensional lattice and isotropic exchange interactions, $J\equiv J_z=J_x$, are assumed. Results for other lattice structures and anisotropic interactions are easily obtained by changing a form factor $\gamma_{\bm{q}}$ in Eq.~(\ref{eq:1}). 

We present in Fig.~\ref{fig:disp} imaginary parts of the dynamical PS susceptibilities for several $J$ and $g$. At $g=0$ [see Fig.~\ref{fig:disp}(a)], a gapless and dispersive low-energy mode exists. This corresponds to the sharp low-energy peak in Fig.~\ref{fig:local}(a), and a purely electronic ``orbiton'' excitation. By introducing a finite $g$ as shown in Fig.~\ref{fig:disp}(b), the high energy multi-peak structure appears, as mentioned in the previous subsection, and the low energy mode remains to be dispersive and gapless. The band width of the low energy mode decreases with increasing $g$ [see Figs.~\ref{fig:disp}(a) and (b)], and increases with increasing $J$  [see Figs.~\ref{fig:disp}(b) and (d)].

Let us focus on the high-energy multi-peak structure. As seen in Figs.~\ref{fig:disp}(c) and (d), spectral distributions show dispersive features for large $J$. Centers of the multi-peak structures are located around $g^2/\omega_0+J$. To clarify nature of the dispersion, we calculate the first and second moments for the high-energy multi-peaks defined by 
$\bar{\omega}^{l}_{\bm{q}}=\means{\omega}_{l \bm{q}}$ and $\Delta \omega^{l}_{\bm{q}}=[\means{\omega^2}_{l \bm{q}}-\means{\omega}^{2}_{l \bm{q}}]^{1/2} $, respectively, where we define 
\begin{align}
\means{f(\omega)}_{l \bm{q}}=
\frac{\int_{\omega_c}^{\infty}d\omega {\rm Im}\chi^{ll}(\bm{q}, \omega) f(\omega)}
{\int_{\omega_c}^{\infty}d\omega {\rm Im}\chi^{ll}(\bm{q}, \omega)} , 
\end{align}
for a function $f(\omega)$ and a cut-off energy $\omega_c$ is chosen to be $\omega_0$. Figures~\ref{fig:disp_width}(a) and (b) show the momentum dependences of $\bar {\omega}_{\bm q}^l$, and ${\bar \omega}_{\bm q}^l \pm \Delta \omega_{\bm q}^l$ for $l=x$ and $z$, respectively. Dispersion in the $xx$-component is larger than that in the $zz$-component. As shown in Eq.~(\ref{eq:1}), the dispersion is almost governed by the transverse PS fluctuation, $v_m^x=\bras{\Phi_0}T^x\kets{\Phi_m}$, for the $xx$ component, and the longitudinal fluctuation, $v_m^z=\bras{\Phi_0}\delta T^z\kets{\Phi_m}$, for the $zz$ component. It is also shown that $\Delta \omega_{\bm q}^l$ is almost proportional to $g$, as seen in Fig.~\ref{fig:disp_width}(c),  where $\Delta \omega^{l}_{\bm{q}}$ are plotted as functions of the JT coupling.

The imaginary parts of the phonon Green's functions are shown in Fig.~\ref{fig:disp_D} for several $J$ and $g$. A flat dispersion at $g=0$ is changed into the gapless dispersive mode by introducing a finite $g$. The low energy mode is identified as a strongly mixed vibronic excitation of orbital and phonon. Spectral intensity is weak in the high energy region above $\omega_0$, where intensive multi-peak structures appear in the orbital channel as shown in Fig.~\ref{fig:disp}.

\section{Low-energy Vibronic Excitation}
\label{sec:low-energy-excit}

In this section, we focus on the low-energy vibronic mode.  

\subsection{Weak Coupling Case}

We assume a small JT coupling, i.e. $g\ll \omega_0,J$, and present a weak coupling formalism based on the perturbational approach. 
The results are compared with the ones obtained in Sec.~\ref{sec:effect-inter-site} and discrepancies between the two are discussed.

We start from the free phonons and orbitons, and introduce the coupling between them. A uniform orbital ordered state for $T^z$ is assumed in the ground state. By applying the Holstein-Primakoff transformation to the PS operators, we have 
$T_i^z=S-b_i^{o\dagger}b_i^o$ and $T_i^x\approx \sqrt{S/2}(b_i^{o\dagger}+b_i^o)$ with $S=1/2$, where $b_i^o$ and $b_i^{o \dagger}$ are the boson operators for ``pure'' electronic orbiton. Hamiltonian in Eq.~(\ref{eq:3}) is rewritten as 
\begin{align}
 {\cal H}^{\rm WC}&=\omega_0\sum_{\bm{q}}
     \left (\tilde{b}_{\bm{q}}^{u\dagger} \tilde{b}_{\bm{q}}^{u}
               +b_{\bm{q}}^{v\dagger} b_{\bm{q}}^{v} \right ) \nonumber \\
 &+\sum_{\bm{q}} \left \{
 \xi_{\bm{q}}b_{\bm{q}}^{o\dagger} b_{\bm{q}}^{o}
      +\frac{1}{2}\zeta_{\bm{k}}(b_{\bm{q}}^{o\dagger} b_{-\bm{q}}^{o\dagger}+h.c.)  \right \} \nonumber\\
&+g\sqrt{\frac{S}{2}}\sum_{\bm{q}}(b_{\bm{q}}^{o\dagger} b_{-\bm{q}}^{v\dagger}+b_{\bm{q}}^{o\dagger} b_{\bm{q}}^v+h.c.)\nonumber\\
&+\frac{g}{\sqrt{N}}\sum_{\bm{k, q}}(b_{\bm{k}}^{o\dagger} b_{\bm{k}+\bm{q}}^o \tilde{b}_{\bm{q}}^u+h.c.),\label{eq:11}
\end{align}
where $b_{\bm{q}}^\eta$ for $\eta=(u, v, o)$ are the Fourier transformations of $b_{i}^\eta$, and $\tilde{b}_{\bm{q}}^{u\dagger}=b_{\bm{q}}^{u\dagger}+\delta_{\bm{q}=0} \sqrt{2N}gS/\omega_0$. We define $\xi_{\bm{q}}=zSJ_z-zSJ_x\gamma_{\bm{q}}/2+g^2S/\omega_0$ and $\zeta_{\bm{q}}=zSJ_x\gamma_{\bm{q}}/2$, and omit constant terms.

By neglecting the fourth line in Eq.~(\ref{eq:11}), which is the higher order in the $1/S$ expansion, the bilinear form is diagonalized by using the Bogoliubov transformation as 
\begin{align}
 {\cal H}^{\rm WC}=\sum_{\bm{q}, \eta=(u, \pm)}{ \Omega}^{\rm WC}_{\bm{q} \eta}\beta_{\bm{q}\eta}^\dagger \beta_{\bm{q}\eta},
 \label{eq:omegasw}
\end{align}
where $\beta_{\bm{q}\eta}$ is the boson operator. The eigen energies are analytically obtained as
\begin{align}
{ \Omega}^{\rm WC}_{\bm{q}u}=\omega_0,
\label{eq:15} 
\end{align}
and 
\begin{align}
{ \Omega}^{\rm WC}_{\bm{q}\pm}&=\Big[ \omega_0^2+\xi_{\bm{q}}^2-\zeta_{\bm{q}}^2\nonumber\\
&\ \ 
\pm\sqrt{(\omega_0^2-\xi_{\bm{q}}^2+\zeta_{\bm{q}}^2)^2+8g^2\omega_0 S(\xi_{\bm{q}}-\zeta_{\bm{q}})} \Big]^{1/2} .
\label{eq:16}
\end{align}
The $u$-phonon does not mix with orbiton in this approximation. 

\begin{figure}[t]
\includegraphics[width=\columnwidth,clip]{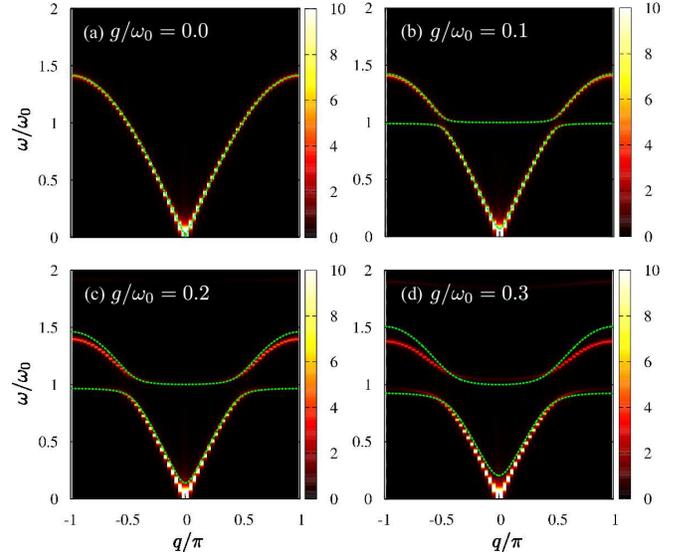}
\caption{(Color online) 
Energy dispersions calculated in the weak coupling approach and contour maps of the imaginary parts of the PS dynamical susceptibilities. Green lines represent ${\Omega}^{\rm WC}_{{\bm q} \eta}$ defined in Eq.~(\ref{eq:omegasw}), and colors maps represent spectral weights $(-1/\pi) {\rm Im} \chi^{xx}({\bm q}, \omega)$. Parameter values are chosen to be (a) $(J/\omega_0, g/\omega_0)=(1,0)$, (b) $(1,0.1)$, (c) $(1, 0.2)$, and (d) $(1,0.3)$. An infinitesimal constant as a damping factor of the spectra is chosen to be $\epsilon/\omega_0=0.1$.
}
\label{fig:SW}
\end{figure}
In Fig.~\ref{fig:SW}, the calculated energy dispersions ${ \Omega}^{\rm WC}_{\bm{q}\pm}$ are compared with $(-1/\pi){\rm Im}\chi^{xx}(\bm{q}, \omega)$ obtained by the method given in Sec.~\ref{sec:formulation-method}. Poles of $(-1/\pi){\rm Im}\chi^{xx}(\bm{q}, \omega)$ are almost reproduced by the weak-coupling results of ${ \Omega}^{\rm WC}_{\bm{q}\pm}$, which are represented as an anti-crossing between the dispersive orbiton and the dispersion-less $v$ phonon. At $g=0$, results obtained by the two methods perfectly coincide with each other. Discrepancies between the two become remarkable with increasing $g$. In particular, the gapless dispersion, required by the Goldstone's theorem, is not reproduced by the weak-coupling approach, in contrast to the method in Secs.~\ref{sec:formulation-method}.
This is due to the fact that, in the weak coupling approach, the two interaction terms between orbiton and phonon in Eq.~(\ref{eq:10}) are not treated on as equal footing: the interaction between $T^z$ and the $u$-phonon is fully considered, but only the lowest order terms of the $1/S$ expansion for the interaction between $T^x$ and the $v$-phonon are taken into account. A similar treatment was adopted in Ref.~\onlinecite{VandenBrink2001}, where the spin wave approximation is applied to Eq.~(\ref{eq:10}) and the fourth line in Eq.~(\ref{eq:11}) is treated by the self-consistent Born approximation. On the other hand, in the present method given in Sec~\ref{sec:formulation-method}, the rotational symmetry in the ${\bm T}$ and ${\bm Q}$ spaces are maintained, and as a result, the gapless mode expected from the Goldstone's theorem appears. 

\subsection{Strong Coupling Case}
\label{sec:sc}

In this subsection, we assume $g \gg \omega_0$, and derive the effective Hamiltonian for the low-energy vibronic state. Calculated results are compared with the results in Sec.~\ref{sec:effect-inter-site}. 

\subsubsection{Low-energy effective Hamiltonian}\label{sec:low-energy-effective}

We focus on the vibronic motion around the potential minima in the lower-adiabatic potential [see Fig.~\ref{fig:adiabatic_plane}(a)], and derive the low-energy effective model from ${\cal H}_i^{\rm MF}$ in the strong coupling limit by following Ref.~\onlinecite{Obrien1964}. In this limit, the conical intersection point shown in Fig.~\ref{fig:adiabatic_plane} is irrelevant, and the Born-Oppenheimer approximation is valid. The vibronic wave-function is given by $\Phi_{n k}(\bm{r},\bm{Q})=\psi_k(\bm{r},\bm{Q})\phi_n^k(\bm{Q})$, where $\psi_k(\bm{r},\bm{Q})$ and $\phi_n^k(\bm{Q})$ are the electronic and lattice wave functions, respectively, and $\bm{r}$ and $\bm{Q}$ are the electron and lattice coordinates, respectively. The adiabatic potentials for ${\cal H}_i^{\rm MF}$ are given as 
\begin{align}
 U^{(k=\pm)}(\bm{Q})=\frac{M\omega_0^2\rho^2}{2}\pm A\rho \pm\frac{1}{2} h_{\rm MF}\cos\theta,
\end{align}
where $\rho=\sqrt{Q_u^2+Q_v^2}$ and $\theta=\tan^{-1}(Q_v/Q_u)$. We assume that $h_{\rm MF}\ll E_{\rm JT}\equiv A^2/(2M\omega_0^2)=g^2/(4\omega_0)$ where $E_{\rm JT}$ is energy gain due to the JT effect. The electronic wave function on the lower adiabatic-potential plane $(k=-)$ up to the order of ${\cal O}(h_{\rm MF})$ is given as 
\begin{align}
\psi_{-}(\bm{r},\bm{Q})&=\psi_{3z^2-r^2}(\bm{r})
\cos\frac{\theta}{2} \left(1
+\frac{h_{{\rm MF}}}{2A\rho}\sin^2\frac{\theta}{2}
 \right)\nonumber\\
&-\psi_{x^2-y^2}(\bm{r})\sin\frac{\theta}{2}
\left(1
-\frac{h_{{\rm MF}}}{2A\rho}\cos^2\frac{\theta}{2}
 \right).
\end{align}
In the case of $h_{\rm MF}=0$, $U^{(-)}$ takes its minima at $\rho=\rho_0\equiv A/(M\omega_0^2)$ for any $\theta$ [see Fig.~\ref{fig:adiabatic_plane}(a)]. In a positive finite $h_{\rm MF}$, this degeneracy is lifted, and $U^{(-)}$ takes its minimum at $\theta=0$. The energy difference between the lower and higher adiabatic planes at $\rho=\rho_0$ in the case of $h_{\rm MF}=0$ is denoted by $2A\rho_0 \equiv g^2/\omega_0 = 4E_{\rm JT}$.

In order to examine the low-energy vibronic excitation, we assume that the zero-point vibration energy ($\omega_0 /2 $) is much smaller than the JT energy gain ($E_{\rm JT}$), and the vibronic motion is confined on the lower adiabatic-potential plane. The effective Hamiltonian for the vibronic motion on this plane is given by 
\begin{align}
& \int d\bm{r}\psi_{-}(\bm{r},\bm{Q})^*
\Bigl[
-\frac{1}{2M}\left(
\frac{\partial^2}{\partial Q_u^2}+\frac{\partial^2}{\partial Q_v^2}
\right)
\nonumber \\
& \hspace{14em}
+U^{(-)}(\bm{Q})
\Bigr] \psi_{-}(\bm{r},\bm{Q}) \nonumber\\
&=\frac{1}{\sqrt{\rho}}\Bigl[
-\frac{1}{2M}\frac{\partial^2}{\partial\rho^2}
+\frac{M\omega_0^2}{2}(\rho-\rho_0)^2-E_{\rm JT}
\Bigr]\sqrt{\rho} \nonumber \\
&-\frac{1}{2M \rho^2}
\frac{\partial^2}{\partial \theta^2} -\frac{h_{\rm MF}}{2} \cos \theta
 -\frac{h_{\rm MF}}{8AM\rho^3} \cos \theta+{\cal O}(h_{\rm MF}^2).\label{eq:6}
\end{align}
The first and second lines of the right hand side in Eq.~(\ref{eq:6}) represent the radial mode and the rotational mode, respectively. Since the characteristic energy for the radial mode, $\omega_0$, is larger than that for the rotational mode, $1/(2M \rho_0^2)=\omega_0^3/g^2$, we focus on the latter at $\rho=\rho_0$. Then the effective Hamiltonian at a single site for the strong coupling limit is given as 
\begin{align}
 {\cal H}_i^{\rm SC}&=-\frac{\omega_0^3}{g^2}\frac{d^2}{d\theta_i^2}-\frac{1}{2}h_{\rm MF}\cos\theta_i.\label{eq:7}
\end{align}
where the last two terms in Eq.~(\ref{eq:6}) is neglected, since $h_{\rm MF}(8AM\rho_0^3)^{-1}=h_{\rm MF}\omega_0^4/(2g^4)$ is much smaller than the kinetic energy $\omega_0^3/g^2$ for rotational mode in the strong coupling case. 

\begin{figure}[t]
\includegraphics[width=\columnwidth,clip]{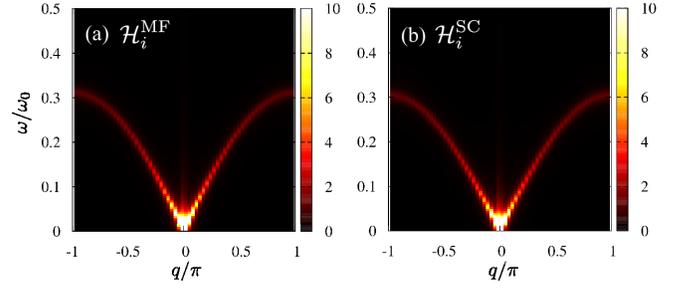}
\caption{(Color online) 
(a) Contour map of imaginary part of the dynamical susceptibilities, and (b) that calculated in the strong coupling approach. Colors represent spectral weights $(-1/\pi) {\rm Im} \chi^{xx}({\bm q}, \omega)$. Parameter values are chosen to be $(J/\omega_0, g/\omega_0)=(1, 4)$. An infinitesimal constant as a damping factor of the spectra is chosen to be $\epsilon/\omega_0=0.1$.
}
\label{fig:low_energy}
\end{figure}
\begin{figure}[t]
\includegraphics[width=\columnwidth,clip]{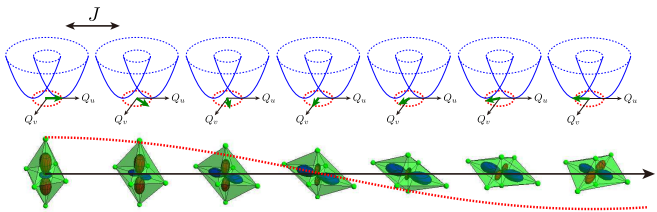}
\caption{(Color online) 
A schematic picture for the collective vibronic mode. Upper panel shows adiabatic-potential planes in a lattice where arrows represent directions of a vector $\bm Q$. Lower panel represents a schematic vibronic state.
}
\label{fig:orbiton}
\end{figure}

The Schr\"odinger equation for the rotational mode, ${\cal H}^{\rm SC}\phi^-_n(\theta)=E_n \phi^-_n (\theta)$, is solved numerically under the anti-periodic boundary condition, $\phi_n(\theta+2\pi)=-\phi_n(\theta)$, required from the condition that $\Phi_{nk}(\bm{r},\bm{Q})$ is single valued. The corresponding vibronic wave function is given by $\Phi_{n-}({\bm r}, \theta)=\psi_-(\bm{r}, \theta)\phi^-_n(\theta)$, and the dynamical orbital susceptibilities are calculated by the method presented in Sec.~\ref{sec:formulation-method}, where ${\cal H}^{\rm MF}_i$ in Eq.~(\ref{eq:5}) is replaced by ${\cal H}^{\rm SC}_i$ in Eq.~(\ref{eq:7}). Results are presented in Fig.~\ref{fig:low_energy}, together with the results obtained in Sec.~\ref{sec:effect-inter-site}. Two results almost coincide with each other. This fact implies that the low-energy excitation mode is identified as the collective vibronic mode where the local rotational mode on the lower adiabatic-potential plane propagates through the inter-site SE interactions. A schematic picture for the collective mode is shown in Fig.~\ref{fig:orbiton}.

\subsubsection{Band width of low-energy collective mode}
\label{sec:LEM}
\begin{figure}[t]
\includegraphics[width=0.8\columnwidth,clip]{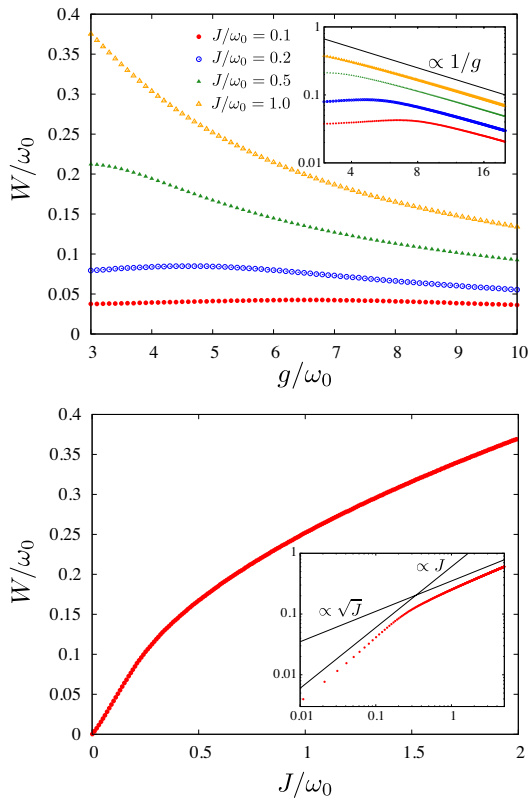}
\caption{(Color online) 
(a) Band widths of the low energy mode as functions of the JT coupling $g$. Inset shows a logarithmic plot. (b) Band width of the low energy mode as a function of the exchange constant $J$. Inset shows a logarithmic plot. A parameter value is chosen to be $g/\omega_0=5$.
}
\label{fig:gap4}
\end{figure}

We focus on the band width of the low-energy excitation mode. As shown in Fig.~\ref{fig:disp}, the band width strongly depends on the JT coupling $g$ [see Figs.~\ref{fig:disp}(a) and (b)], as well as the exchange constant $J$ [see Figs.~\ref{fig:disp}(b) and (d)]. Detailed analyses are given in Figs.~\ref{fig:gap4}(a) and (b), where the band width defined by $W \equiv \Omega_{q=\pi, \eta_0}$, where $\eta_0$ indicates the lowest branch, are plotted as a function of $g$ and $J$, respectively.
The band width is renormalized as $1/g$ for large $J$ and $g$, and is almost independent of $g$ for small $J$ and $g$. As for the $J$ dependence, $W$ is almost proportional to $J$ for small $J$ and is proportional to $\sqrt{J}$ for large $J$. 

These results are interpreted from the Hamiltonian in Eq.~(\ref{eq:7}) by the perturbational schemes as follows. Since the band width, i.e. the excitation energy at the zone boundary, corresponds to the orbital excitation energy at a single site under the mean-field. This is equivalent to the energy difference $\Delta$ between the ground state and the first excited state in ${\cal H}_i^{\rm SC}$. 
In the weak SE interaction or the weak JT coupling $[h_{\rm MF}(\sim J) \ll 2\omega_0^3/g^2]$, where the second term in ${\cal H}^{\rm SC}_i$ is treated as a perturbational term, we have $\Delta=h_{\rm MF}(\bras{\Phi_u}T^z\kets{\Phi_u}-\bras{\Phi_v}T^z\kets{\Phi_v})=h_{\rm MF}/2$, where $\kets{\Phi_u}$ and $\kets{\Phi_v}$ are the degenerate ground states in the case of $h_{\rm MF}=0$. A factor $1/2$ originates from reductions of the matrix elements, $\bras{\Phi_u}T^z\kets{\Phi_u}=-\bras{\Phi_v}T^z\kets{\Phi_v}=1/4$, known as the Ham's reduction effect.~\cite{Ham1968} This result explains that $W$ is proportional to $J$ and is almost independent of $g$. 
On the other side, in the strong SE interaction or the strong JT coupling [$h_{\rm MF} (\sim J) \gg 2\omega_0^3/g^2$], a deep potential minimum exists at $\theta=0$, and the Hamiltonian in Eq.~(\ref{eq:7}) is expanded by $\theta$ as 
\begin{align}
 {\cal H}^{\textrm{SC}}=-\frac{\omega_0^3}{g^2}\frac{d^2}{d \theta^2}+\frac{h_{\rm MF}}{4}\theta^2-\frac{h_{\rm MF}}{48}\theta^4 
, \label{eq:8}
\end{align}
where the constant terms are omitted. By taking the last term as a perturbation, 
we have $\Delta=\sqrt{h_{\rm MF}}\omega_0^{3/2}/g-\omega_0^3/(4g^2) $, which explains that $W$ is proportional to $\sqrt{J}$ and $1/g$. 

\section{Comparison between the $E\otimes e$ and $E\otimes b_1$ systems}\label{sec:comp-betw-cases}

To clarify characteristics of the low-energy vibronic excitation in the $E \otimes e$ JT system, we introduce a system where the doubly degenerate $e_g$ orbitals couple to a non-degenerate vibrational mode, and compare the two results. The interaction between the orbital and phonon in this system is given by 
\begin{align}
 {\cal H}_i^{ E \otimes b_1 }=\omega_0 b_i^{u\dagger}b_i^{u}-gT_i^z(b_i^{u\dagger}+b_i^u) , 
\label{eq:18}
\end{align}
where one phonon mode couples to the electronic orbital. A similar Hamiltonian was studied in Ref.~\onlinecite{Schmidt2007}. We analyze the Hamiltonian in Eq.~(\ref{eq:hmf}), where ${\cal H}_i^{\rm JT}$ is replaced by ${\cal H}_i^{E \otimes b_1}$, by using the method presented in Sec.~\ref{sec:formulation-method}. A uniform orbital order for $T^z$ is assumed in the ground state. 

\begin{figure}[t]
\includegraphics[width=\columnwidth,clip]{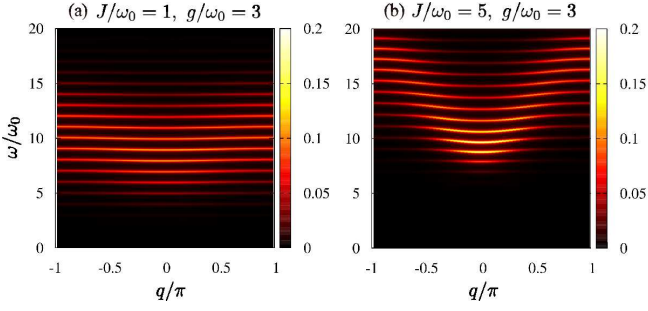}
\caption{(Color online) 
Contour maps of imaginary parts of the dynamical susceptibilities in the $E \otimes b_1$ system. Colors represent spectral weights $(-1/\pi) {\rm Im} \chi^{xx}({\bm q}, \omega)$. Parameter values are chosen to be (a) $(J/\omega_0, g/\omega_0)=(1, 3)$ and (b) $(5,3)$. An infinitesimal constant as a damping factor of the spectra is chosen to be $\epsilon/\omega_0=0.1$.
}
\label{fig:holstein}
\end{figure}
\begin{figure}[t]
\includegraphics[width=0.8\columnwidth,clip]{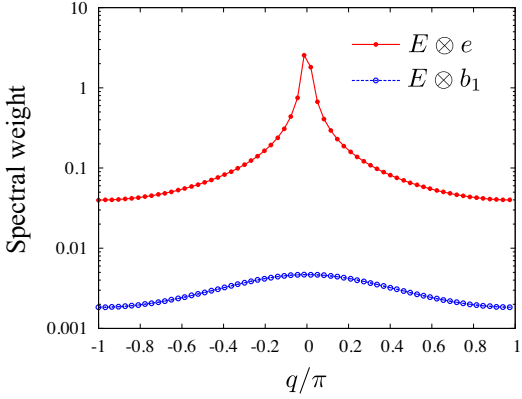}
\caption{(Color online) 
Momentum dependences of the spectral weights $-(1/\pi){\rm Im}\chi^{xx}({\bf q}, \omega)$ at $\omega=\Omega_{q\eta_0}$. Red line with filled circles is for the $E\otimes e$ system and blue line with open circles is for  the $E\otimes b_1$ system. Parameter values are chosen to be $(J/\omega_0, g/\omega_0)=(1, 3)$. 
}
\label{fig:intensity}
\end{figure}
The imaginary parts of the dynamical susceptibility $(-1/\pi){\rm Im}\chi^{xx}({\bm q}, \omega)$ are presented in Fig.~\ref{fig:holstein}. Momentum dependent high-energy multi-peaks appear and a center of the multi-peaks is located around $g^2/\omega_0+J$. These characteristics are similar to the high-energy excitations in the $E\otimes e$ JT system presented in Sec.~\ref{sec:effect-inter-site} and are attributed to the excitations from the lower to higher adiabatic-potential planes. On the other hand, intensity of the low energy mode is much weaker than that in the $E \otimes e$ system shown in Fig.~\ref{fig:disp}. Detailed comparison for the low energy mode is shown in Fig.~\ref{fig:intensity} where the momentum dependences of the spectral weight $-(1/\pi){\rm Im}\chi_{q}^{xx}({q}, \omega)$ at poles are presented. Spectral intensities in the $E\otimes b_1$ system are almost one order smaller than those in the $E\otimes e$ system. 

This difference is attributed to a dimensionality in the lattice-coordination space. In the $E\otimes b_1$ system, the adiabatic potential is defined in the one-dimensional $Q_u$ coordinate and shows a double-well type potential where minima exist at two discrete values, as shown in Fig.~\ref{fig:adiabatic_plane}(b). The excitation inside the lower-adiabatic plane is an Ising-type. With increasing $g$, distance between the coordinates where the potential takes the minima increases, 
and an overlap between the wave-functions at two minima is reduced. As a result, amplitude in the low-energy excitation mode is weakened. 
This is in contrast qualitatively to the $E \otimes e$ system where the adiabatic-potential planes are defined in the two-dimensional $Q_u$-$Q_v$ plane. There is a continuous degeneracy for the potential minima in the lower adiabatic-potential plane, as shown in Fig.~\ref{fig:adiabatic_plane}(a), and the Bloch-wave type vibration wave-function are extended along the potential minima. As a result, the gapless low-energy vibronic excitation exists even in the strong coupling regime.

\section{Discussion and Summary}
\label{sec:discussion-summary}

In this section, we discuss effects of several factors which are neglected so far: the anharmonic lattice potential, the cooperative lattice effect, the anisotropic exchange interactions, and the spin degree of freedom, all of which are able to be introduced in the present formalisms. 

\begin{figure}[t]
\includegraphics[width=0.8\columnwidth,clip]{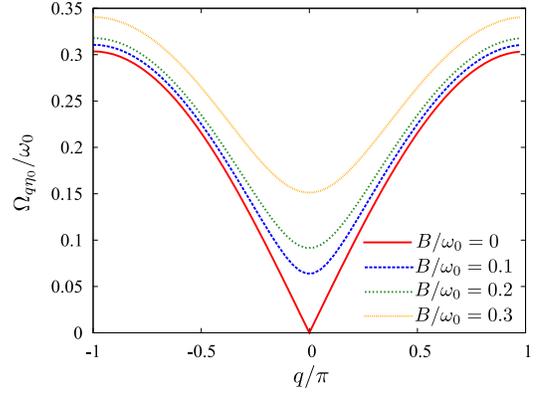}
\caption{(Color online) 
Energy dispersions of the low-energy mode for several anharmonic lattice potentials $(B)$. Parameter values are chosen to be $(J/\omega_0, g/\omega_0)=(1, 4)$. 
}
\label{fig:anharmonicp}
\end{figure}

First, we show effects of the anharmonic lattice potential. This is known to play key roles on orbital orders in real materials. The anharmonic lattice potential is treated within the strong coupling approach presented in Sect.~\ref{sec:sc}. The local Hamiltonian corresponding to Eq.~(\ref{eq:7}) is given by 
\begin{align}
  {\cal H}_i^{\textrm{AH}}=-\frac{\omega_0^3}{g^2}\frac{d^2}{d\theta_i^2}-\frac{1}{2}h_{\rm MF}\cos\theta_i-B\cos 3\theta_i,\label{eq:17}
\end{align}
where the third term represents the anharmonic lattice potential with a positive constant $B$. This term stabilizes the JT distortions at $\theta=0$ and $\pm 2\pi/3$. 
The Schr\"odinger equation for ${\cal H}_i^{\textrm{AH}}$ is solved numerically, and the dynamical PS susceptibilities are calculated by the method in Sec.~\ref{sec:formulation-method}, where ${\cal H}^{\rm MF}_i$ in Eq.~(\ref{eq:5}) is replaced by ${\cal H}^{\textrm{AH}}_i$. Results of the dispersion relation of the low-energy collective excitations are presented in Fig.~\ref{fig:anharmonicp}. The excitation gap opens by introducing finite $B$, because the rotational symmetry in the lower adiabatic-potential plane is broken by the anharmonic potential. 

\begin{figure}[t]
\includegraphics[width=1.0\columnwidth,clip]{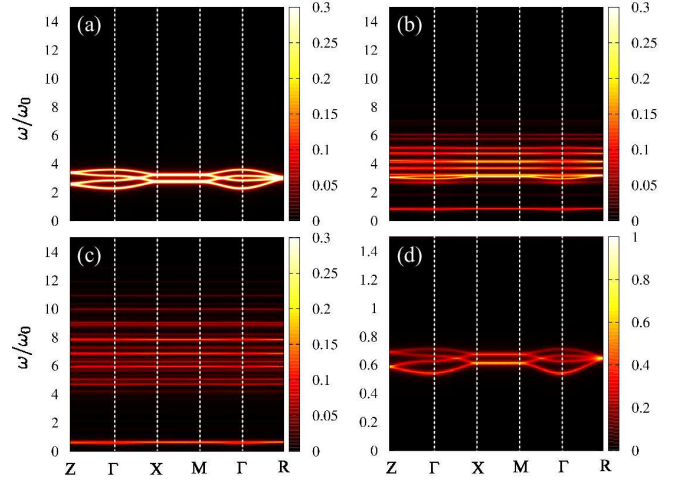}
\caption{(Color online) 
Imaginary parts of the PS susceptibilities at (a) $g/\omega_0=0$, (b) 1, and (c) 2. (d) shows an expansion of the low-energy region of (c). Parameter values are chosen to be $(J_1/\omega_0, J_2/\omega_0)=(1, 0.5)$. An infinitesimal constant as a damping factor of the spectra is chosen to be $\epsilon/\omega_0=0.05$ in (a)-(c) and $0.01$ in (d).
} 
\label{fig:realistic}
\end{figure}

An excitation energy gap is also opened by the anisotropic SE interactions. So far, we assume $J_x=J_z$ which ensures the continuous symmetry in the ${\bm T}$ and ${\bm Q}$ spaces. In realistic materials, however, the SE interactions are anisotropic; $J_x \ne J_z$, and $T^x_iT^z_j$ as well as $T_i^x+T^x_j$ terms exist. The anisotropic SE interactions are able to be dealt with in the present formalism. We demonstrate this application in a Kugel-Khomskii type Hamiltonian~\cite{Kugel1972,Kugel1974,Ishihara1997} where the doubly degenerate $e_g$ orbitals are introduced at each site in a simple cubic lattice. The Hamiltonian is given by~\cite{Ishihara2000} 
\begin{align}
{\cal H}_{\rm KK}&= -2J_1\sum_{\means{ij}_\mu}\left(\frac{1}{4}+\bm{S}_i\cdot\bm{S}_j\right)\left(\frac{1}{4}-\tau_i^\mu\tau_j^\mu\right)\nonumber\\
&\ \ \ -2J_2
\sum_{\means{ij}_\mu}\left(\frac{1}{4}-\bm{S}_i\cdot\bm{S}_j\right)\left(\frac{3}{4}+\tau_i^\mu\tau_j^\mu+\tau_i^\mu+\tau_j^\mu\right),
\label{eq:KK}
\end{align}
where $J_1$ and $J_2$ are the exchange constants, $\tau_i^\mu$ is the bond-dependent pseudo-spin operator defined by $\tau_i^\mu=\cos(2\pi n_\mu/3)T_i^z-\sin(2\pi n_\mu/3)T_i^x$ with $(n_z,n_x,n_y)=(0,1,2)$, a subscript $\mu(=x, y, z)$ indicates a direction of the $ij$ bond, and ${\bm S}_i$ is the spin operator. 
This model has been studied for the orbital structures and excitations in LaMnO$_3$ and KCuF$_3$.~\cite{Ishihara2000}
We focus on the orbital dynamics and neglect spin excitation, which will be discussed later. We apply the present method given in Sec.~\ref{sec:formulation-method}, where ${\cal H}_J$ is replaced by ${\cal H}_{\rm KK}$ with above approximation. The PS dynamical susceptibility is defined by
\begin{align}
 \chi_{\Lambda\Lambda'}^{ll'}(\bm{q},\omega)=-i\int_{0}^\infty dt\brasd{0}[\tilde T_{-\bm{q}\Lambda}^l(t),\tilde T_{\bm{q}\Lambda'}^{l'}]\ketsd{0}e^{i\omega t-\epsilon t},
\end{align}
for $l=z,x$, where $\Lambda(=A,B,C,D)$ describes the four orbital sublattices. 
Figures~\ref{fig:realistic}(a)-(d) show $(-1/\pi){\rm Im}\chi_{AA}^{xx}({\bm q}, \omega)$. At $g=0$ [see Fig.~\ref{fig:realistic}(a)], the pure ``orbiton'' shows gapful excitation around $J$ due to the anisotropic exchange interactions.~\cite{Ishihara2000} Four modes are attributed to the four sublattices. By introducing the JT coupling $g$ [see Fig.~\ref{fig:realistic}(b) and (c)], the pure orbiton modes are changed into the low-energy modes and the high-energy multi-peaks. It is worth noting that, even in the strong coupling cases, characteristic dispersion relations, which are similar to the pure orbiton excitations, appear in the low-energy collective mode, as shown in Fig.~\ref{fig:realistic}(d). This dispersive low energy modes is interpreted as a vibronic collective modes where intensity and energy are strongly renormalized. 

The cooperative JT (CJT) effect, neglected so far, plays sometime essential roles on the orbital order as well as the excitation dispersions.~\cite{Kanamori1960,Kaplan1995} This interaction is attributed to the interaction between the lattice displacement in different JT centers, and is able to be treated in the same way with the inter-site SE interaction in the present formalism. 
Let us consider, for an example, a simple model for the interaction between the NN JT centers as $\sum_{\langle ij \rangle ll'} K_{ll'} Q_{il} Q_{jl'} $ with the spring constants $K_{ll'}$. The MF decoupling is introduced as 
$Q_{il} Q_{jl'} =
          Q_{il}  \langle Q_{jl'} \rangle + 
          \langle Q_{il} \rangle  Q_{jl'}  -\langle Q_{il} \rangle \langle Q_{jl'} \rangle + \delta Q_{il}\delta Q_{jl'}$, where $\delta Q_{il}=Q_{il} -\means{Q_{il}}$. 
Hamiltonian in Eq.~(\ref{eq:5}) is replaced by 
\begin{align}
 {\cal H}&=-\sum_{\means{ij}}(J_z \delta T_i^z \delta T_j^z+J_x T_i^x T_j^x) \\
&+\sum_{\means{ij} lm} K_{ll'} \delta Q_{il} \delta Q_{jl'}
 +\sum_i {\cal H}_i^{\rm MF}.
\end{align}
The on-site term 
\begin{align}
{\cal H}_i^{\rm MF}=-h_{\rm MF}T_i^z +h_{Q} Q_{iu} +{\cal H}_i^{\rm JT}, 
\end{align}
where $h_Q$ is a mean-field acting on $Q_{iu}$, is diagonalized, as explained in Sec.~\ref{sec:formulation-method}. Both $h_{\rm Q}$ and $h_{\rm MF}$ are calculated self-consistently, and $\delta Q_{il}$ is represented by the boson operators, $a_{im}$ and $a_{im}^\dagger$, in a similar way to Eq.~(\ref{eq:tx}). Then, the effects of CJT are taken into account on an equal footing to the SE interaction. 

We touch briefly the spin degree of freedom. As shown in the Kugel-Khomskii type model in Eq.~(\ref{eq:KK}), the NN interactions are expressed by products of the spin part and the orbital part. This interactions can be treated in the same way of ${\cal H}_{J}$ in Eq.~(\ref{eq:5}) as follow. The SE Hamiltonian are represented by a sum of the spin interactions, the orbital interactions and the spin-orbital interactions. The MF decouplings are applied to each term such as 
${\bm S}_i \cdot {\bm S}_j \rightarrow 
          {\bm S}_i \cdot \langle {\bm S}_j \rangle + 
          \langle {\bm S}_i \rangle \cdot {\bm S}_j  -\langle {\bm S}_i \rangle \cdot \langle {\bm S}_j \rangle$, 
$T_i^l T_j^m \rightarrow 
          T_i^l  \langle T_j^m \rangle + 
          \langle T_i^l \rangle  T_j^m  -\langle T_i^l \rangle \langle T_j^m \rangle$, 
$T_i^l {\bm S}_i \cdot {\bm S}_j T_j^m \rightarrow 
          T_i^l {\bm S}_i \cdot \langle {\bm S}_j T_j^m\rangle + 
          \langle T_i^l {\bm S}_i \rangle \cdot {\bm S}_j T_j^m  -\langle T_i^l {\bm S}_i \rangle \cdot \langle {\bm S}_j T_j^m \rangle$ 
and so on. Then, the local Hamiltonian ${\cal H}^{\rm MF}_i$ corresponding to Eq.~(\ref{eq:5}) is numerically solved in the spin, orbital and phonon Hilbert spaces. Three-kinds of the boson operators are required to be introduced: orbiton, magnon and orbiton-magnon which changes spin and orbital states at the same site, simultaneously.~\cite{Cyrot1975,VandenBrink1998,Sikora2006} The bilinear boson Hamiltonian obtained by the generalized spin-wave approximation is diagonalized by the Bogoliubov transformation, and the diagonal Hamiltonian such as Eq.~(\ref{eq:omega}) is obtained. 
It is expected that, in a collinear spin ordered state as a ground state, finite mixings occur between magnon and orbiton-magnon, and between orbiton and phonon due to the conservation of the spin angular momentum. In a non-collinear spin ordered state, four kind excitations are mixed with each other, and the JT effect affect the spin dynamics directly. 
The present method is also valid for systems where the relativistic spin-orbit interaction is relevant. Intra-atomic magnetic structures with the spin-orbit interaction are able to be taken into account in the similar way to the JT coupling. The calculated spin-orbital mixed excitations due to the spin-orbit interactions will be compared with experimental observations in the $4d$ and $5d$ transition-metal compounds, such as iridium oxides, as well as the $3d$ transition-metal oxides. 

Finally, we discuss vibronic excitations from the view point of experimental observations. Present calculations, for example the results in Fig.~\ref{fig:realistic}, are directly applicable to the excitation dynamics in the several $e_g$ orbital ordered systems, e.g. LaMnO$_3$, KCuF$_3$ and others. Detailed calculations for each material will be presented in future works. Here, we suggest that experimental observations rather depend on magnitudes of the JT coupling. In a weak JT coupling regime, dispersive excitations, being similar to the pure electronic orbitons, are expected. Weak multiple structures due to the JT coupling will appear in the orbiton bands. The excitation energy is characterized by the SE interactions which are of the order of 10-100meV in typical transition-metal compounds. Such vibronic excitations can be detected by the resonant x-ray scattering experiments. On the other hand, in a strong JT coupling regime, excitation energies of the dispersive branches are renormalized and shift to lower than the bare JT phonon frequency. This is the energy range for the optical spectroscopy measurements. The recently developed non-resonant inelastic x-ray scattering technique is applicable to detect the dispersions of the renormalized vibronic modes. The inelastic neutron scattering, which directly accesses to the phonon channel [see Fig.~\ref{fig:disp_D}], is another candidate to detect the dispersive collective vibronic excitations. In the strong coupling regime, observation of the characteristic momentum dependent intensities/energies of the high-energy multi-peak structures 
also provides several information for orbital excitation and JT coupling.

%%%%%%%%%%%%%%%%%%%%
%%   Conclusion   %%
%%%%%%%%%%%%%%%%%%%%

In conclusion, we present a theoretical framework of vibronic excitations where both the local vibronic excitations and the inter-site orbital interaction are taken into account on an equal footing. We confirm that the present formalism is valid from the weak to strong coupling regimes. Two kinds of excitations are identified; the low-energy collective vibronic mode connected to orbiton, and the high-energy multi-peaks originating from the single JT center. The present formalism is applicable to a wide range of correlated electron models with the orbital degrees of freedom.

\begin{acknowledgments}
One of the authors (JN) thanks S.~Yamazaki and Y. Yamaji for helpful discussions. This work was supported by KAKENHI from MEXT and Tohoku University ``Evolution'' program. JN is supported by the global COE program ``Weaving Science Web beyond Particle-Matter Hierarchy'' of MEXT, Japan. Parts of the numerical calculations are performed in the supercomputing systems in ISSP,  the University of Tokyo. 
\end{acknowledgments}

\appendix

\section{Relation to Random Phase Approximation}
\label{sec:equiv-with-rand}

In this Appendix, we derive another expression, a RPA-type expression, for the dynamical PS susceptibility which is given in Eqs.~(\ref{eq:19}) and (\ref{eq:20}). Since both of the expressions in Eq.~(\ref{eq:19}) with Eq.~(\ref{eq:20}), and Eq.~(\ref{eq:21}) are derived from the same Hamiltonian without any approximations, the two expressions are equivalent.

We start from ${\cal H}={\cal H}_J+{\cal H}_{\rm JT}$ where ${\cal H}_{\rm JT}$ is defined in Eq.~(\ref{eq:2}), and assume a general form for the SE interactions as 
\begin{align}
 {\cal H}_J=-\sum_{\means{ij}}\sum_{ll'=x,z}J_{ll'} T_i^l T_j^{l'},
\end{align}
where $J_{l l'}$ represent the SE interactions between $T_i^l$ and $T_j^{l'}$. We assume the uniform orbital order of $\langle T_z\rangle$ in the mean-field ground state of the Hamiltonian ${\cal H}$. Hamiltonian corresponding to Eq.~(\ref{eq:5}) is given by 
\begin{align}
 {\cal H}=-\sum_{\means{ij}ll'}J_{ll'} \tilde{T}_i^l \tilde{T}_j^{l'}
+\sum_i {\cal H}_i^{\rm MF},
\end{align}
with 
\begin{align}
{\cal H}_i^{\rm MF}=-h_{\rm MF}T_i^z +{\cal H}_i^{\rm JT}.
\end{align} 
We define $\tilde{T}_i^z=T_i^z-\means{T^z}$, and $\tilde{T}_i^x=T_i^x$. By introducing the generalized Holstein-Primakoff transformation in the same way with Eqs.~(\ref{eq:tx}) and~(\ref{eq:tz}), Hamiltonian is written by the boson operators as 
\begin{align}
{\cal H}&=\sum_{\bm{q}}\sum_{mn}\bigl[(\Delta E_n \delta_{mn}-z \gamma_{\bm{q}}\sum_{ll'}J_{ll'} v_m^l  v_n^{l'})a_{\bm{q}m}^\dagger a_{\bm{q}n}\nonumber\\
 &-\frac{z\gamma_{\bm{q}}}{2} \sum_{ll'}J_{ll'}v_m^l v_n^{l'}(a_{\bm{q}m}^\dagger a_{-\bm{q}n}^\dagger+h.c)\bigr],\label{eq:4}
\end{align}
where definitions of the symbols are the same as those in Eq.~(\ref{eq:1}).

We consider the propagator for the boson operator as 
\begin{align}
 P_{mn}(\bm{q},\tau)&=-\means{T_{\tau}\phi_{-\bm{q}m}(\tau)\phi_{\bm{q}n}}, 
\end{align}
where $\phi_{\bm{q}n}=a_{\bm{q}n}+a_{-\bm{q}n}^\dagger$. For convenience, we present the Matsubara formalism in finite temperature. The Fourier transformation of the propagator is given by  
\begin{align}
 P_{mn}(\bm{q},i\omega_p)&=\int_{0}^\beta d\tau P_{mn}(\bm{q},\tau)e^{i\omega_p \tau},
\end{align}
where $T_{\tau}$ is the time-ordering operator, $\omega_p$ is the Matsubara frequency, $\means{\cdots}$ represents the thermal average and ${\cal O}(\tau)=e^{\tau {\cal H}}{\cal O}e^{-\tau {\cal H}}$. The orbital susceptibility is given as $\chi^{ll'}(\bm{q},i\omega_p)=\sum_{mn}v_m^l v_n^{l'} P_{mn}(\bm{q},i\omega_p)$. 

The equation of motion of the propagator is obtained by 
\begin{align}
 i\omega_p P_{mn}=\Delta E_m Q_{mn},\label{eq:13}
\end{align}
where
\begin{align}
 Q_{mn}(\bm{q},i\omega_p)&=-\int_{0}^\beta d\tau\means{T_{\tau}\pi_{-\bm{q}m}(\tau)\phi_{\bm{q}n}}e^{i\omega_p \tau},
\end{align}
with $\pi_{\bm{q}n}=a_{\bm{q}n}-a_{-\bm{q}n}^\dagger$ and $\Delta E_{m}=E_{m}-E_0$. The equation of motion of the propagator $Q_{mn}$ is also obtained as 
\begin{align}
 i\omega_p Q_{mn}&=2\delta_{mn} +\Delta E_m P_{mn}\nonumber\\
&\ \ \ \ -2z\gamma_{\bm{q}}\sum_{ll'}J_{ll'}(\hat{M}^{l'l} \hat{P})_{mn} . 
\label{eq:14}
\end{align}
We define $\hat{M}^{ll'}=\bm{v}^{l}\otimes\bm{v}^{l'}$ where $v_n^l=\bras{n}\tilde{T}^l\kets{0}$. 
From Eqs.~(\ref{eq:13}) and~(\ref{eq:14}), we have   
\begin{align}
\hat{P}(\bm{q},i\omega_p)=\left[\hat{P}_0(i\omega_p)-z\gamma_{\bm{q}} \sum_{ll'}J_{ll'}\hat{M}^{l'l}\right]^{-1},
\end{align}
with $[\hat{P}_0(i\omega_p)]_{mn}=\delta_{mn}\{(i\omega_p)^2/(2\Delta E_n)-\Delta E_n/2\}$.  
The susceptibility is obtained as
\begin{align}
 \chi^{ll'}(\bm{q},i\omega_p)&={\rm Tr}\left[\left(\hat{P}_0^{-1}+z\gamma_{\bm{q}}\sum_{kk'}J_{kk'} \hat{M}^{k'k}\right)^{-1}\hat{M}^{l'l}\right], \nonumber \\
&={\rm Tr}\left[\hat{\chi}_{\rm loc}^{ll'}\left(1+z\gamma_{\bm{q}} \sum_{kk'}J_{kk'}\hat{\chi}_{\rm loc}^{kk'}\right)^{-1}\right] , 
\label{eq:21}
\end{align}
where $\hat{\chi}_{\rm loc}^{ll'}=\hat{P}_0 \hat{M}^{l'l}$. 
Finally, we have a RPA-type expression as 
\begin{align}
 \chi^{ll'}(\bm{q},i\omega_p)=\left[\bar{\chi}_{\rm loc}(i\omega_p)\left(1+\bar{J}(\bm{q})\bar{\chi}_{\rm loc}(i\omega_p)\right)^{-1}\right]_{ll'},
\end{align}
where we define 
$[{\bar \chi}_{\rm loc}(i \omega_p)]_{ll'}={\rm Tr}[\hat{\chi}_{\rm loc}^{ll'}(i \omega_p)]$ and 
$[\bar{J}(\bm{q})]_{l l'}=z\gamma_{\bm{q}}J_{ll'}$.

%*****************************************************************************
\noindent
$\ast$Present address: Department of Applied Physics, University of Tokyo, Tokyo, 113-8656, Japan

% \bibliographystyle{apsrev}
% \bibliography{../../library}

\begin{thebibliography}{99} 
%\begin{references}

\bibitem{Tokura2000}
Y. Tokura and N. Nagaosa, Science {\bf 288}, 462 (2000).

\bibitem{Maekawa2004}
 S.~Maekawa, T.~Tohyama, S.~E.~Barnes, S.~Ishihara, W.~Koshibae, and G.~Khaliullin, {\it Physics of Transition Metal Oxides} (Springer Verlag, Berlin, 2004).

%-----
%optics
\bibitem{Saitoh2001}
E.~Saitoh, S.~Okamoto, K.~T.~Takahashi, K.~Tobe, K.~Yamamoto, T.~Kimura, S.~Ishihara, S.~Maekawa, and Y.~Tokura, 
\journal{Nature}{410}{180}{2001}.

\bibitem{Miyasaka2005}
S.~Miyasaka, S.~Onoda, Y.~Okimoto, J.~Fujioka, M.~Iwama, N.~Nagaosa, and Y.~Tokura
\journal{\PRL}{94}{076405}{2005}.

\bibitem{Ulrich2006}
C.~Ulrich, A.~G\"ossling, M.~Gr\"uninger, M.~Guennou, H.~Roth, M.~Cwik, T.~Lorenz, G.~Khaliullin, and B.~Keimer,
\journal{\PRL}{97}{157401}{2006}.

\bibitem{Sugai2006}
S.~Sugai, A.~Kikuchi, and Y.~Mori
\journal{\PRB}{73}{161101}{2006}.

\bibitem{Benckiser2008}
E.~Benckiser, R.~R\"uckamp, T.~M\"oller, T.~Taetz, A.~M\"oller, A.~A.~Nugroho, T.~T.~M.~Palstra, G.~S.~Uhrig, and M.~Gr\"uninger,
\journal{New~J.~Phys.}{10}{053027}{2008}.

%x-ray
\bibitem{Inami2003}
T.~Inami, T.~Fukuda, J.~Mizuki, S.~Ishihara, H.~Kondo, H.~Nakao, T.~Matsumura, K.~Hirota, Y.~Murakami, S.~Maekawa, and Y.~Endoh,
\journal{\PRB}{67}{045108}{2003}.

\bibitem{Tanaka2004}
Y.~Tanaka, A.~Q.~R.~Baron. Y.-J.~Kim, K.~J.~Thomas, J.~P.~Hill, Z.~Honda, F.~Iga, S.~Tsutui, D.~Ishihkawa and C.~S.~Nelson, 
New.~J.~Phys. {\bf 6} 161 (2004). 

\bibitem{Ulrich2009}
C. Ulrich, L. J. P. Ament, G. Ghiringhelli, L. Braicovich, M. M. Sala, N. Pezzotta, T. Schmitt, G. Khaliullin, J. van den Brink, H. Roth, T. Lorenz, and B. Keimer,  
\journal{\PRL}{103}{107205}{2009}.

\bibitem{Ishii2011}
K.~Ishii, S.~Ishihara, Y.~Murakami, K.~Ikeuchi, K.~Kuzushita, T.~Inami, K.~Ohwada, M.~Yoshida, I.~Jarrige, N.~Tatami, S.~Niioka, D.~Bizen, Y.~Ando, J.~Mizuki, S.~Maekawa, and Y.~Endoh,
\journal{\PRB}{83}{241101}{2011}.

\bibitem{Schlappa2012}
J.~Schlappa, K.~Wohlfeld, K.~J.~Zhou, M.~Mourigal, M.~W.~Haverkort, V.~N.~Strocov, L.~Hozoi, C.~Monney, S.~Nishimoto, S.~Singh, A.~Revcolevschi, J.-S.~Caux, L.~Patthey, H.~M.~R{\o}nnow, J.~van~den~Brink, and T.~Schmitt,
\journal{Nature}{485}{82}{2012}.


%observation theory
\bibitem{Inoue1997}
J.~Inoue, S.~Okamoto, S.~Ishihara, W.~Koshibae, Y.~Kawamura, and S.~Maekawa,
Physica~B {\bf 237-238}, 51 (1997). 

%\bibitem{Okamoto2002}
%S.~Okamoto, S.~Ishihara, and S.~Maekawa,
%\journal{\PRB}{66}{014435}{2002}.

\bibitem{Ishihara2004}
S.~Ishihara,
\journal{\PRB}{69}{075118}{2004}.

\bibitem{Ishihara2005}
S.~Ishihara, Y.~Murakami, T.~Inami, K.~Ishii, J.~Mizuki, K.~Hirota, S.~Maekawa and Y.~Endoh
New~J.~Phys. {\bf 7}, 119 (2005). 

\bibitem{Haverkort2010}
M.~W.~Haverkort,
\journal{\PRL}{105}{167404}{2010}.

\bibitem{Ament2011}
L.~J.~P.~Ament, G.~Khaliullin, and J.~van~den~Brink,
\journal{\PRB}{84}{020403}{2011}.



%-------

\bibitem{Cyrot1975}
M.~Cyrot and C.~Lyon-Caen, 
J.~Phys.~Paris {\bf 36}, 253 (1975).

\bibitem{Komarov1975}
A.~G.~Komarov, L.~I.~Korovin, and E.~K.~Kudinov, 
Sov.~Phys.~Sol.~Stat. {\bf 17}, 1531 (1976). 

\bibitem{Ishihara1996}
S.~Ishihara, J.~Inoue and S.~Maekawa, 
Physica C {\bf 263}, (1996).

\bibitem{Ishihara1997}
S.~Ishihara, J.~Inoue and S.~Maekawa, 
Phys. Rev. B {\bf 55}, 8280 (1997).

\bibitem{Ishihara2000}
S.~Ishihara, and S.~Maekawa, 
Phys. Rev. B {\bf 62}, 2338 (2000).
%JT --> pressure

\bibitem{Bala2000}
J.~Ba{\l}a and A.~M.~Ole\'s,
\journal{\PRB}{62}{R6085}{2000}.

\bibitem{VandenBrink2001}
J.~van~den~Brink,
\journal{\PRL}{87}{217202}{2001}.

\bibitem{Gehring1975}
G.~A.~Gehring, and K.~A.~Gehring, 
Rep. Prog. Phys. {\bf 38}, 1 (1975). 

\bibitem{Kaplan1995}
M.~D.~Kaplan and B.~G.~Vekhter, {\it Cooperative Phenomena in Jahn-Teller Crystals} (Plenum Press, New York, 1995) 
Cahpter 6.

\bibitem{Bersuker2006}
As a review, 
I. B. Bersuker, 
{\it The Jan-Teller Effect} 
(Cambridge University Press, Cambridge, 2006)
Chapter 6. 

\bibitem{Allen1999}
P.~B.~Allen and V.~Perebeinos,
\journal{\PRL}{83}{4828}{1999}.

%-----
%\bibitem{Han2003}
%J.~Han, O.~Gunnarsson, and V.~Crespi,
%\journal{\PRL}{90}{167006}{2003}.
%
%\bibitem{Hotta2006}
%T.~Hotta,
%\journal{\PRL}{96}{197201}{2006}.
%
%\bibitem{Nasu2012b}
%J.~Nasu and S.~Ishihara,
%arXiv:1209.0239.

%-----KK

\bibitem{Kugel1972}
K.~I.~Kugel and D.~I.~Khomskii,
\journal{Sov.~Phys.~Lett.}{15}{446}{1972}.

\bibitem{Kugel1974}
K.~I.~Kugel and D.~I.~Khomskii,
\journal{Sov.~Phys.~-JETP}{37}{725}{1974}.

\bibitem{Sarfatt1964}
J. Sarfatt, and A. M. Stoneham, 
Proc. Phys. Soc. {\bf 91}, 214 (1964). 


% --- generalized SW

\bibitem{Papanicolaou1988}
N.~Papanicolaou,
\journal{Nucl.~Phys.~B}{305}{367}{1988}.

\bibitem{Onufrieva1985}
F.~P.~Onufrieva,
\journal{Sov.~Phys.~JETP}{89}{2270}{1985}.

\bibitem{Shiina2003}
R.~Shiina, H.~Shiba, P.~Thalmeier, A.~Takahashi, and O.~Sakai,
\journal{\JPSJ}{72}{1216}{2003}.

\bibitem{Kusunose2001}
H.~Kusunose and Y.~Kuramoto
\journal{\JPSJ}{70}{3076}{2001}.

\bibitem{Joshi1998}
A.~Joshi, M.~Ma, F.~Mila, D.~N.~Shi, and F.~C.~Zhang,
\journal{\PRB}{60}{6584}{1999}.

\bibitem{Colpa1978}
J.~H.~P.~Colpa,
\journal{Physica~A}{93}{327}{1978}.
%------

\bibitem{Koller2004}
W.~Koller, D.~Meyer, and A.~C.~Hewson,
\journal{\PRB}{70}{155103}{2004}.

\bibitem{Obrien1964}
M.~C.~M.~O'Brien,
\journal{Proc.~Roy.~Soc.~(London)}{A281}{323}{1964}.

\bibitem{Ham1968}
F.~Ham,
\journal{\PR}{166}{307}{1968}.


%------
%orbiton phonon
\bibitem{Schmidt2007}
K.~P.~Schmidt, M.~Gr\"uninger, and G.~S.~Uhrig,
\journal{\PRB}{76}{075108}{2007}.

%------


\bibitem{Kanamori1960}
J.~Kanamori,
\journal{\JAP}{31}{S14}{1960}.

%spin-orbital
\bibitem{VandenBrink1998}
J.~van~den~Brink, W.~Stekelenburg, D.~I.~Khomskii, G.~A.~Sawatzky, and K.~I.~Kugel,
\journal{\PRB}{58}{10276}{1998}.

\bibitem{Sikora2006}
O.~Sikora and A.~M.~Ole\'s,
\journal{Physica~Status~Solidi~(B)}{243}{133}{2006}.

%\end{references}
\end{thebibliography}

%*****************************************************************************

\end{document}